\documentclass[12pt]{article}
\usepackage{epsf}
\usepackage{amsmath,amssymb}
\usepackage{latexsym}
\usepackage{graphicx,color}
\usepackage{wrapfig}
\usepackage{latexsym}
\usepackage{graphics}

\usepackage{color}
\usepackage{delarray,color}
\usepackage{fancybox}  
\newcommand {\beq}{\begin{eqnarray}}
\newcommand {\eeq}{\end{eqnarray}}

\usepackage{tabularx}

\catcode`\@=11
\@addtoreset{equation}{section}

\catcode`@=12
\relax

\newcommand{\siki}[1]{(\ref{#1})}

\begin{document}
\baselineskip 0.7cm

\begin{titlepage}

\setcounter{page}{0}

\renewcommand{\thefootnote}{\fnsymbol{footnote}}

\begin{flushright}
YITP-08-55\\
\end{flushright}

\vskip 1.35cm

\begin{center}
{\Large \bf 
Orbifolding the Membrane Action
}

\vskip 1.2cm 

{\normalsize
Seiji Terashima$^1$\footnote{terasima(at)yukawa.kyoto-u.ac.jp} and Futoshi Yagi$^1$\footnote{futoshi(at)yukawa.kyoto-u.ac.jp}
}

\vskip 0.8cm

{ \it
$^1$Yukawa Institute for Theoretical Physics, Kyoto University, Kyoto 606-8502, Japan
}

\end{center}

\vspace{12mm}

\centerline{{\bf Abstract}}

We study a simple class of orbifolds 
of the ${\cal N}=6$ Chern-Simons Matter theory 
proposed by Aharony, Bergman,
Jafferis and Maldacena. 
They are considered 
as world volume theories of membranes 
probing $\mathbb{C}^4/(\mathbb{Z}_k \times \mathbb{Z}_n)$
and include new membrane theories 
with ${\cal N}=4$ supersymmetries.
We find that the moduli spaces of them are
consistent with the fact that 
they probe $\mathbb{C}^4/
(\mathbb{Z}_k \times \mathbb{Z}_n)$.

\end{titlepage}
\newpage

\section{Introduction}

The works of Bagger and Lambert \cite{BL1,BL2,BL3}
and Gustavsson \cite{G}
made breakthrough on the study of    
multiple membranes.
Recently, 
three dimensional ${\cal N}=6$ Chern-Simons theory with gauge group $U(N) \times U(N)$ 
were constructed by Aharony, Bergman,
Jafferis and Maldacena
(ABJM) \cite{ABJM} and many aspects of the 
theory have been explored recently
\cite{Klebanov}-\cite{ABJMr2}.\footnote{
It has been shown that 
the Bagger-Lambert theory with $SO(4)$ structure
can be written as a Chern-Simons-matter theory \cite{vR}.
and it describes membranes on an orbifold \cite{LT, Mfold}.}
This theory at level $k$ is suggested to describe the world volume theory 
of M2 branes probing $\mathbb{C}^4/\mathbb{Z}_k$,
where the discrete group $\mathbb{Z}_k$ act complex coordinate $y^A$ $(A=1,\cdots,4)$ as 
\begin{eqnarray}
\left( y^1 , y^2 , y^3 , y^4 \right) \to 
\left( e^{\frac{2\pi i}{k}} y^1 , \,\, e^{\frac{2\pi i}{k}} y^2, \,\, 
e^{\frac{2\pi i}{k}} y^3 , \,\, e^{\frac{2\pi i}{k}} y^4 \right) .
\end{eqnarray}
This proposal is very interesting and 
it is desirable to investigate this theory further.
In order to get more examples of 
the multiple membrane theory, 
the orbifolding will be very useful.
Indeed, 
the orbifold of  
the ABJM theory with gauge group $SU(2) \times SU(2)$,
i.e. the BLG case, and its moduli space were studied 
in \cite{Fuji}
and two different $\mathbb{Z}_n$ orbifolds 
of the ABJM theory with $U(N) \times U(N)$ were studied in
\cite{Klebanov}.\footnote{
Our models considered in section 4.1 and section 4.2 were 
the same models considered in \cite{Imamura}, 
whose analysis turns out to be consistent with ours
although we independently did this work.}

In this paper, 
we study a general class of 
$\mathbb{Z}_n$ orbifold theories 
of ABJM with $U(N) \times U(N)$ gauge group,
and find another ${\cal N}=4$ supersymmetric 
membrane theory.
This discrete group $\mathbb{Z}_n$ act as rotation of the phase of 
the complex coordinate $y^A$ as
$$\left( y^1 , y^2 , y^3 , y^4 \right) \to 
\left( e^{\frac{2\pi i}{n_1}} y^1 , \,\, e^{\frac{2\pi i}{n_2}} y^2, \,\, 
e^{-\frac{2\pi i}{n_3}} y^3 , \,\, e^{-\frac{2\pi i}{n_4}} y^4 \right).$$
We also study the moduli spaces of the $\mathbb{Z}_n$
orbifolds of the ABJM action for following two cases 
\begin{eqnarray}
&{\rm (I)}& \qquad (n_1,n_2,n_3,n_4)=(n,n,-n,-n), \cr
&{\rm (II)}& \qquad (n_1,n_2,n_3,n_4)=(n,n,\infty, \infty), 
\end{eqnarray}
which preserve ${\cal N}=4$ supersymmetry.
For convenience, we allow $n_A=\infty$, which means that
$y^A$ is invariant under this action. 
We find they are consistent
with the fact that 
they probe $\mathbb{C}^4/
(\mathbb{Z}_k \times \mathbb{Z}_n)$,
where we assume that 
$k=k' \, n$ where $k' \in \mathbf Z$.

The organization of this paper is as follows.
We begin in section 2 with a brief summary of ABJM theory.
Next we discuss in section 3 the classification of $\mathbb{Z}_n$ orbifold of ABJM theory
and analyze supersymmetry of the orbifold theory.  
In section 4, we analyze the moduli space
of several examples of orbifold gauge theory.
Section 5 is devoted to conclusions and discussions.

\section{ABJM theory}
In this section, we briefly review the ABJM theory \cite{ABJM}.
This theory is $U(N)_1 \times U(N)_2$ gauge theory with corresponding vector superfields 
${\cal V}, \hat{\cal V}$, which are not dynamical, and with 
hypermultiplet superfields in bifundamental and anti-bifundamental representation
\begin{eqnarray}
{\cal Z}^A = Z^A + \theta \zeta^A + \theta^2 F^A \cr
{\cal W}^A = W^A + \theta \omega^A + \theta^2 G^A,
\label{superfield}
\end{eqnarray}
where $A=1,2$. 
The global symmetries 
which can be seen explicitly are $SU(2)_1$ which act only on  
$\cal Z$ and $SU(2)_2$ which act on $\cal W$.
The matter contents and the manifest symmetries 
are summarized in Table \ref{matter_contents}.

\begin{table}[ht]
\centering
\begin{tabular}{|c|c|c|c|c|c|c|}
& $U(N)_1$ & $U(N)_2$ & $SU(2)_1$ & $SU(2)_2$ \cr
\hline
\vspace{-4mm}
&&&& \cr
$\cal Z$ & $N$ & $\bar{N}$ & ${\bf 2}$ & ${\bf 1}$ \cr
$\cal W$ & $\bar{N}$ & $N$ & ${\bf 1}$ & ${\bf 2}$ \cr 
$\cal V$ & Adjoint & ${\bf 1}$ & ${\bf 1}$ & ${\bf 1}$ \cr
$\cal \hat{V}$ & {\bf 1} &  Adjoint & ${\bf 1}$ & ${\bf 1}$ \cr 
\end{tabular}
\caption{Matter contents and symmetry}
\label{matter_contents}
\end{table}

Actually, this theory also has $SU(2)_R$ symmetry, under which
$(Z^1,W^{\dagger 1})$ and $(Z^2,W^{\dagger 2})$
form multiplets.
As this symmetry does not commute with 
$SU(2)_1\times SU(2)_2$ symmetry mentioned above,
the theory must have $SU(4)_R$ symmetry as a whole.
Four scalars 
\begin{eqnarray}
y^A \equiv \{ Z^A , W^{\dagger A} \} \label{def_y}
\end{eqnarray}
form a fundamental representation of this $SU(4)_R$
while  four fermions 
\begin{eqnarray}
\psi_A \equiv 
\{ \varepsilon_{AB}\zeta^B e^{\frac{i \pi}{4}} ,  
\varepsilon_{AB} \omega^{\dagger B} e^{\frac{- i \pi}{4}} \}
\label{def_psi}
\end{eqnarray}
form an anti-fundamental representation \cite{ABJM,Klebanov}. 
Existence of $SU(4)_R \simeq Spin(6)_R$ symmetry indicates ${\cal N}=6$ supersymmetry.
Six supersymmetry generators are in the vector representation of the $Spin(6)_R$,
or equivalently, in the antisymmetric tensor representation of the $SU(4)_R$.

The action of this theory in a superspace formalism is given by \cite{Klebanov}.
After integrating out auxiliary fields, the action is given by \cite{ABJM,Klebanov}
\begin{eqnarray}
S=\int d^3 x \left[ \frac{k}{4 \pi} \varepsilon^{\mu\nu\lambda} \mathrm{Tr} \left(
A_{\mu} \partial _{\nu} A_\lambda + \frac{2i}{3} A_{\mu} A_{\nu} A_{\lambda} 
- \hat{A}_{\mu} \partial_{\nu} \hat{A}_{\lambda} 
- \frac{2i}{3} \hat{A}_{\mu} \hat{A}_{\nu} \hat{A}_{\lambda} \right) \right. \cr
- \mathrm{Tr} (D_{\mu}Z)^{\dagger} D^{\mu}Z - \mathrm{Tr} (D_{\mu}W)^{\dagger} D^{\mu}W
+ i \mathrm{Tr} \zeta^{\dagger} \gamma^{\mu} D_{\mu} \zeta \cr
+ i \mathrm{Tr} \, \omega^{\dagger} \gamma^{\mu} D_{\mu} \omega 
 - V_{\mathrm{bos}} - V_{\mathrm{ferm}} \Big]
\end{eqnarray}
with the potential $V_{\mathrm{bos}}=V^{\mathrm{bos}}_D+V^{\mathrm{bos}}_F$, where
\begin{eqnarray}
&& \!\! \!\! \!\! \!\! \!\! \!\! \!\! \!\! \!\! \!\! \!\! \!\! \!\! \!\!
V^{\mathrm{bos}}_D = \frac{4 \pi^2}{k^2} \mathrm{Tr} \left[ 
 (Z^A Z_A^{\dagger} + W^{\dagger A} W_A) 
 (Z^B Z_B^{\dagger} - W^{\dagger B} W_B )
 (Z^C Z_C^{\dagger} - W^{\dagger C} W_C ) \right. \cr \vspace{1mm}
&& \qquad \quad + ( Z_A^{\dagger} Z^A + W_A W^{\dagger A} ) 
 (Z_B^{\dagger} Z^B - W_B W^{\dagger B} )
 (Z_C^{\dagger} Z^C - W_C W^{\dagger C} ) \cr 
&& \qquad \quad -2 Z_A^{\dagger} (Z^B Z_B^{\dagger} - W^{\dagger B} W_B ) 
 Z^A (Z_C^{\dagger} Z^C - W_C W^{\dagger C} ) \cr
&& \left. \qquad \quad
- 2 W^{\dagger A} (Z_B^{\dagger} Z^B - W_B W^{\dagger B} )
 W_A (Z^C Z_C^{\dagger} - W^{\dagger C} W_C )
\right] \cr
&& \!\! \!\! \!\! \!\! \!\! \!\! \!\! \!\! \!\! \!\! \!\! \!\! \!\! \!\!
V^{\mathrm{bos}}_F = - \frac{16\pi^2}{k^2} \mathrm{Tr} \left[ 
W^{\dagger A} Z_{B}^{\dagger} W^{\dagger C} W_A Z^B W_C 
-  W^{\dagger A} Z_{B}^{\dagger} W^{\dagger C} W_C Z^B W_A \right. \cr
&&\left. \qquad \quad
+ Z_{A}^{\dagger} W^{\dagger B} Z_{C}^{\dagger} Z^A W_B Z^C
-  W^{\dagger A} Z_{B}^{\dagger} W^{\dagger C} Z^C W_B Z^A \right]
\end{eqnarray}
and $V_{\mathrm{ferm}}=V^{\mathrm{ferm}}_D + V^{\mathrm{ferm}}_F$, where
\begin{eqnarray}
&& \!\! \!\! \!\! \!\! \!\!
V^{\mathrm{ferm}}_D = \frac{2\pi i}{k} \mathrm{Tr} \left[ 
\left( \zeta^{\dagger}_A \zeta^A - \omega_A \omega^{\dagger A} \right)
\left( Z_B^{\dagger} Z^B - W_B W^{\dagger B} \right) \right. \cr
&&  \qquad  \qquad  \qquad  \qquad
- \left. \left( \zeta^A \zeta^{\dagger}_A - \omega^{\dagger A} \omega_A \right)
\left( Z^B Z_B^{\dagger} - W^{\dagger B} W_B \right) 
 \right] \cr
&& \qquad + \frac{8\pi i}{k} \mathrm{Tr} \left[ 
\left( Z^{\dagger}_A \zeta^A - \omega_A W^{\dagger A} \right)
\left( \zeta_B^{\dagger} Z^B - W_B \omega^{\dagger B} \right) \right. \cr
&&   \qquad  \qquad  \qquad  \qquad
\left. - \left( \zeta^A Z^{\dagger}_A - W^{\dagger A} \omega_A \right)
\left( Z^B \zeta_B^{\dagger} - \omega^{\dagger B} W_B \right) 
 \right] \cr
&&  \!\! \!\! \!\! \!\! \!\!
V^{\mathrm{ferm}}_F = \frac{2\pi}{k} \varepsilon_{AC} \varepsilon^{BD}
\mathrm{Tr} \left[ 2 \zeta^A W_B Z^C \omega_D + 2 \zeta^A \omega_B Z^C W_D \right.
 \left. + Z^A \omega_B Z^C W_D + \zeta^A W_B \zeta^C W_D
\right] \cr
&& \qquad + \frac{2\pi}{k} \varepsilon_{AC} \varepsilon^{BD}
\mathrm{Tr} \left[ 2 \zeta^{\dagger}_A W^{\dagger B} Z^{\dagger}_C \omega^{\dagger D} 
+ 2 \zeta_A^{\dagger} \omega^{\dagger B} Z^{\dagger}_C W^{\dagger D} \right. \cr
&& \qquad\qquad\qquad\qquad\qquad\qquad
 \left. + Z^{\dagger}_A \omega^{\dagger B} Z^{\dagger}_C W^{\dagger D} 
+ \zeta^{\dagger}_A W^{\dagger B} \zeta^{\dagger}_C W^{\dagger D} 
\right] 
\end{eqnarray}
Here, $A^{\mu}$ and $\hat{A}^{\mu}$ are gauge fields for $U(N)_1$ and $U(N)_2$, respectively. 

Full global symmetry acting on the moduli space of this theory is actually $SU(4)_R \times U(1)_b$.
The $U(1)_b$ symmetry is originally an over all $U(1)$ part 
of the gauge symmetry $U(N)_1 \times U(N)_2$,
whose corresponding gauge field is 
$\mathrm{Tr} \left( A^{\mu} - \hat{A}^{\mu} \right)$.
Due to the Chern-Simons term, this $U(1)_b$ symmetry is ungauged on the moduli space 
except for its subgroup $\mathbb{Z}_k$ \cite{ABJM},
where $\mathbb{Z}_k$ acts on 4 complex coordinates $y^A$ as
\beq
y^A \to e^{2\pi i / k} y^A \qquad A=1,2,3,4 \label{Z_vec}.
\eeq

This theory is suggested to describe the world volume theory of $N$ M2 branes 
probing $\mathbb{C}^4/\mathbb{Z}_k$.
In particular, for $k=1$ and $k=2$, the global symmetry is expected to be $Spin(8)_R$,
which indicate ${\cal N}=8$ supersymmetry.
The global $SU(4)_R \times U(1)_b$ symmetry can be regarded as a subgroup 
of the full $Spin(8)_R$ symmetry.

\section{Orbifold action and supersymmetry}

In this section, we consider orbifold gauge theory of the ABJM theory 
and calculate how many supersymmetries are preserved.
First, we introduce eight supercharges, which are transformed as a spinor
of the full $Spin(8)_R$ symmetry. 
They are labeled by spin weight $s_A=\pm 1/2$.
From chirality condition, we have the condition
\beq
s_1 + s_2 + s_3 + s_4 = 2 \mathbb{Z}. \label{chirality}
\eeq
The $\mathbb{Z}_k$ acts on the SUSY generators as 
\beq
Q_{s_1 , \cdots s_4} \to e^{(s_1 + s_2 + s_3 + s_4) 2\pi i/k } Q_{s_1 , \cdots s_4}.
\label{Q_transf}
\eeq
Thus, the spinors left invariant by this orbifold action satisfies the condition 
\beq 
s_1+s_2+s_3+s_4 = k \mathbb{Z}. \label{cond1}
\eeq
For $k=1,2$, this condition are included by \siki{chirality}.
The theory has ${\cal N} =8$ supersymmetry in this case.
For $k \ge 3$, six out of eight spinors satisfies this condition,
which indicates ${\cal N} =6$ supersymmetry.

We consider further orbifolding by the following $\mathbb{Z}_n$ action 
\beq
y^A \to e^{2\pi i /n_A} y^A, \qquad A=1,2,3,4 \label{vec}
\eeq
which is a subgroup of the global $SU(4)_R \times U(1)_b$ symmetry.  
For convenience, we allow 
$n_A=\infty$, which means that $y^A$ is invariant under this action.
As this transformation acts on the SUSY generators as
\beq
Q_{s_1 , \cdots s_4} \to 
\exp \left[ 2\pi i \left( \frac{s_1}{n_1} + \frac{s_2}{n_2} 
 + \frac{s_3}{n_3} + \frac{s_4}{n_4} \right)  \right] 
Q_{s_1 , \cdots s_4}, \label{susy_gen}
\eeq
the invariant generators under this action satisfy
\beq
\frac{s_1}{n_1} + \frac{s_2}{n_2} + \frac{s_3}{n_3} + \frac{s_4}{n_4} = \mathbb{Z}
\label{cond2}.
\eeq
Solution of \siki{cond1} and \siki{cond2} are the remaining supersymmetry generators
under a given $\mathbb{Z}_n$ orbifold action \siki{vec}.
Generically, if you find the solution satisfying \siki{cond1} and \siki{cond2},
same $\{ n_A \}$ with opposite sign of $\{ s_A \}$ of the solution is also a solution. 
Thus, the number of the remaining supersymmetry generators is always even.

The above discussion is based on the assumed $Spin(8)_R$ symmetry,
which is not manifest in the ABJM action even for $k=1,2$.
However, the same results will be obtained even if we start from the 
global $SU(4)_R \times U(1)_b$ symmetry,
which is manifest in the action given in \cite{Klebanov}.
The important point here is that $U(1)_b$ transformation commutes 
with supersymmetry transformation.
Indeed, we can see it from the explicit ${\cal N}=6$ transformation
\cite{Gaiotto,Hosomichi2,BL4,Terashima},
taking account that $U(1)_b$ act on scalars and fermions as
$y^A \rightarrow e^{i \phi} y^A$ and 
$\psi^{\dagger A} \rightarrow e^{-i \phi} \psi^{\dagger A}$.
Thus, only the $\mathbb{Z}_n$ orbifold action included in the $SU(4)_R$ part
can break the supersymmetry.
As the six spinors above, which satisfy \siki{cond1}, are in an
anti-symmetric tensor representation of this $SU(4)_R$ symmetry,
we see that spinors invariant under $\mathbb{Z}_n \subset SU(4)_R$ 
remain as symmetries of the orbifolded theory.

In section 4, we will use the ${\cal N}=2 $ superfields
instead of the $SU(4)_R$ manifest form
because we translate the orbifold action to
the orbifolds of the superfields
for the examples we will study.

For simplicity, we assume that $k \ge 3$.
In the following, we analyze which $\mathbb{Z}_n$ orbifold action preserves 
supersymmetry and how many supersymmetries are preserved under that action.
As $n_A=\infty$ and $n_A= 1$ are special,
we classify the solutions according to the number of them.
Note that the action \siki{vec} for scalars is the same for $n_A=1$ as that for $n_A=\infty$.
However, the action for SUSY generators \siki{susy_gen} is different between $|n_A|=1$ and $n_A=\infty$.
This difference come from whether the orbifold action 
include the $\mathbb{Z}_2 (\subset Spin(8)_R)$ action
which changes the sign of matter fermions while does not act on scalars. 
As we will see later, the number of the preserving supersymmetry is different 
between $n_A=\infty$ and $n_A=1$ for some cases.
On the other hand, for example, $n_A=1$ and $n_A=-1$ are equivalent both for \siki{Z_vec} and \siki{susy_gen}.

\subsection*{(I) $\mathbb{C}^4$}

We consider the case where all the $n_A$ are either $\infty$ or $1$.
This orbifold action does not act on the scalars $y^A$ at all,
but potentially acts on SUSY generators as $\mathbb{Z}_2$ action.

\subsubsection*{(i) All the $n_A$ are $\infty$}
This case is trivial because it corresponds to the unorbifolded case.

\subsubsection*{(ii) Three of $n_A$ are $\infty$ and one of $n_A$ is $1$}
Let $n_1=1$ and $n_2=n_3=n_4=\infty$ without loss of generality.
The condition \siki{cond2} reduces to
$$s_1=\mathbb{Z}.$$
This cannot be satisfied.

\subsubsection*{(iii) Two of $n_A$ are $\infty$ and two of $n_A$ are $1$}
From \siki{Q_transf}, we see that $\mathbb{Z}_n$ acts trivially not only on
the scalars but also the SUSY generators. 
Thus, this case also corresponds to the unorbifolded case.

\subsubsection*{(iv) One of $n_A$ is $\infty$ and three of $n_A$ are $1$}
Let $n_1=n_2=n_3=1$ and $n_4=\infty$ without loss of generality.
The condition \siki{cond2} reduces to
$$s_1 + s_2 + s_3 = \mathbb{Z}.$$
This condition cannot be satisfied.

\subsubsection*{(v) All the $n_A$ are $1$}
This case also corresponds to the unorbifolded case.


\subsection*{(II) $\mathbb{C}^3 \times \mathbb{C}/\mathbb{Z}_n$}
We consider the case where three out of four $n_A$s are either $\infty$ or $1$, the other is generic.
This is the case where $\mathbb{Z}_n$ acts on only one out of the four scalars.

\subsubsection*{(i) Three of $n_A$ are $\infty$}
Let $n_2=n_3=n_4=\infty$ without loss of generality.
The condition \siki{cond2} reduces to
$$\frac{s_1}{n_1}=\mathbb{Z}.$$
This cannot be satisfied for any finite $n_1$.

\subsubsection*{(ii) Two of $n_A$ are $\infty$, one of $n_A$ is $1$}
Let $1<|n_1|<\infty$, $n_2= 1$, $n_3=n_4=\infty$ without loss of generality.
The condition \siki{cond2} reduces to
$$\frac{s_1}{n_1} + s_2 = \mathbb{Z}.$$
This cannot be satisfied because $1<|n_1|<\infty$.

\subsubsection*{(iii) One of $n_A$ is $\infty$, two of $n_A$ are $1$}
Let $1<|n_1|<\infty$, $n_2=n_3=1$, $n_4=\infty$ without loss of generality.
The condition \siki{cond2} reduces to
$$\frac{s_1}{n_1} + s_2 + s_3 = \mathbb{Z}.$$
This cannot be satisfied because $1<|n_1|<\infty$.

\subsubsection*{(iv) Three of $n_A$ are $1$}
Let $1<|n_1|<\infty$, $n_2=n_3=n_4=1$ without loss of generality.
The condition \siki{cond2} reduces to
$$\frac{s_1}{n_1} + s_2 + s_3 + s_4 = \mathbb{Z}.$$
This cannot be satisfied because $1<|n_1|<\infty$.

\subsection*{(III) $\mathbb{C}^2 \times \mathbb{C}^2/\mathbb{Z}_n$}
We consider the case where two out of four $n_A$s are either $\infty$ or $1$, others are generic.
This class of orbifolding has been investigated in \cite{Klebanov}.

\subsubsection*{(i) Two of $n_A$ are $\infty$}
Let $1<|n_1|,|n_2|<\infty$, $n_3=n_4=\infty$.
The condition \siki{cond2} reduces to
$$\frac{s_1}{n_1}+\frac{s_2}{n_2}=\mathbb{Z}.$$
As $s_A=\pm 1/2$, in order for this condition to have solutions,
$(n_1,n_2)$ have to satisfy
$$\frac{1}{n_1} \pm \frac{1}{n_2} = 2\mathbb{Z},$$
which is equivalent as 
$$n_1 = \pm n_2 .$$
We consider the cases for each sign in the following.

\subsubsection*{(1) $n_1=n_2\equiv n$}

As $|n| \ge 2$ and $k \ge 3$, \siki{cond1} and \siki{cond2} are reduced to the following two equations
\beq
s_1+s_2=0 , \qquad s_3+s_4=0. \label{1st_ex}
\eeq
Four spinors 
$$
(s_1,s_2,s_3,s_4)=(+,-,+,-), \,\, (+,-,-,+), \,\, (-,+,+,-), \,\, (-,+,-,+)
$$
satisfies these conditions, which indicates ${\cal N}=4$ supersymmetry.
For brevity, we wrote $``+''$ for $1/2$ and $``-''$ for $-1/2$.
This is the ``non-chiral orbifold gauge theories'' investigated by \cite{Klebanov}.

\subsubsection*{(2) $n_1=-n_2\equiv n$}
In this case, \siki{cond1} and \siki{cond2} are reduced to the following two equations
\beq
s_1-s_2=0 , \qquad 2s_1+s_3+s_4=0. \label{2nd_ex}
\eeq
Two spinors 
\beq
(s_1,s_2,s_3,s_4)=(+,+,-,-), \,\, (-,-,+,+)
\eeq
satisfies these conditions, which indicates ${\cal N}=2$ supersymmetry.
This is the ``chiral orbifold gauge theories'' investigated by \cite{Klebanov}.

\subsubsection*{(ii) One of $n_A$ is $\infty$, another of $n_A$ is $1$, others are generic}
Let $1<|n_1|,|n_2|<\infty$, $n_3=1$, $n_4=\infty$ without loss of generality.
The condition \siki{cond2} reduces to
\beq
\frac{s_1}{n_1}+\frac{s_2}{n_2} + s_3=\mathbb{Z}. \label{2_2_1}
\eeq
As $s_A=\pm 1/2$, in order for this condition to have solutions,
$(n_1,n_2)$ have to satisfy
$$\frac{1}{n_1} \pm \frac{1}{n_2} = 2\mathbb{Z}+1,$$
which is equivalent as 
$$|n_1| = |n_2| = 2.$$

When $n_1$ and $n_2$ have the same sign, \siki{2_2_1} further reduce to
$$ s_1 + s_2 + 2s_3 = 2\mathbb{Z},$$
which is equivalent as 
$$ s_1 + s_2 = 2\mathbb{Z} + 1.$$
Together with \siki{cond1}, the remaining spinors are following two
\beq
(s_1, s_2, s_3, s_4) = (+,+,-,-), \,\, (-,-,+,+),
\eeq
which indicates ${\cal N}=2$ SUSY.
When $n_1$ and $n_2$ have the opposite sign, the condition becomes
$$ s_1 - s_2 = 2\mathbb{Z} + 1.$$
In this case, the remaining spinors are
\beq
(s_1, s_2, s_3, s_4) = (+,-,-,+), \,\, (+,-,+,-), \,\, (-,+,-,+), \,\, (-,+,+,-),
\eeq
which indicates ${\cal N}=4$ SUSY.

Compared with the results in (III)-(i), in which two of $n_A$ are $\infty$,
the preserving supersymmetry is exchanged between $n_1=n_2$ and $n_1=-n_2$.

\subsubsection*{(iii) Two of $n_A$ are $1$, others are generic}
This orbifold action is equivalent as (III)-(i).

\subsection*{(IV) $\mathbb{C} \times \mathbb{C}^3/\mathbb{Z}_n$}
We consider the case where one out of four $n_A$s is $\infty$ or $1$.

\subsubsection*{(i) One of $n_A$ is $\infty$.}
Let $n_4=\infty$.
The condition \siki{cond2} reduces to
$$\frac{s_1}{n_1}+\frac{s_2}{n_2}+\frac{s_3}{n_3}=\mathbb{Z}.$$
From \siki{cond1}, two of the $s_A$ are $1/2$ while others are $-1/2$. 
Thus, in order for this condition to have solutions,
$(n_1,n_2,n_3)$ have to satisfy
$\frac{1}{n_1} + \frac{1}{n_2} - \frac{1}{n_3} = 2\mathbb{Z}.$
As $|n_1|,|n_2|,|n_3| \ge 2$, the absolute value of the left hand is less than two.
Thus, the condition we should solve is
$$\frac{1}{n_1} + \frac{1}{n_2} - \frac{1}{n_3} = 0.$$ 
This equation has infinitely many solutions as 
$$(n_1,n_2,n_3) = (2n,2n,n), \,\, (6n,3n,2n), \,\, (12n,4n,3n) \cdots .$$
For all these cases, the remaining spinors are following two
\beq
(s_1, s_2, s_3, s_4) = (+,+,-,-), \,\, (-,-,+,+),
\eeq
which indicates ${\cal N}=2$ supersymmetry.

\subsection*{(V) $\mathbb{C}^4/\mathbb{Z}_n$}

We consider the case where all the $n_A$ are generic.

First, we seek for solutions preserving ${\cal N}=6$ supersymmetry.
As discussed above, two of the $s_A$ are $1/2$ while others are $-1/2$ from \siki{cond1},
and it makes six combinations.
Thus, the solution for $\{ s_A \}$ should be the following six
\beq
(s_1,s_2,s_3,s_4)=(+,+,-,-), \,\, (-,-,+,+), \,\, (+,-,+,-), \,\, (-,+,-,+), \,\, 
\cr(-,+,+,-), \,\, (+,-,-,+) \nonumber
\eeq
In order to for these six to be all solutions,
$\{ n_A \}$ must satisfy following three conditions,
\beq
\frac{1}{n_1}+\frac{1}{n_2}-\frac{1}{n_3}-\frac{1}{n_4} = 2\mathbb{Z}\cr 
\frac{1}{n_1}-\frac{1}{n_2}+\frac{1}{n_3}-\frac{1}{n_4} = 2\mathbb{Z}\cr 
\frac{1}{n_1}-\frac{1}{n_2}-\frac{1}{n_3}+\frac{1}{n_4} = 2\mathbb{Z}
\eeq
The solution of these equations are 
$$(n_1,n_2,n_3,n_4) = (n,n,n,n).$$
Though $(2,2,-2,-2)$ is also a solution, it is equivalent to $(2,2,2,2)$.

Next, we seek for solutions preserving ${\cal N}=4$ supersymmetry.
As discussed above, two of the $s_A$ are $1/2$ while others are $-1/2$ from \siki{cond1}.
First, let $(s_1, s_2, s_3, s_4) = (+,-,+,-)$ be a solution.
Then, $(s_1, s_2, s_3, s_4) = (-,+,-,+)$ is also a solution.
In order to preserve ${\cal N}=4$ supersymmetry, one more pair should be a solution,
which we suppose $(+,-,-,+)$ and $(-,+,+,-)$ without loss of generality.
Then, the conditions which $\{ n_A \}$ should satisfy are
\beq
\frac{1}{n_1}-\frac{1}{n_2}+\frac{1}{n_3}-\frac{1}{n_4} = 2\mathbb{Z} , \qquad
\frac{1}{n_1}-\frac{1}{n_2}-\frac{1}{n_3}+\frac{1}{n_4} = 2\mathbb{Z}
\eeq
The solution of these equations are only
$$(n_1,n_2,n_3,n_4) = (n,n,m,m).$$
Especially important example is $(n,n,-n,-n)$, 
which we will investigate in detail in the following section.

There are plenty of orbifold action preserving ${\cal N}=2$ supersymmetry.
One of the most important example is 
$$(n_1,n_2,n_3,n_4)=(n,n,n,3n),$$
which is discussed for $n=1$ in the last section of \cite{Klebanov}.

\subsection*{Summary}

Here, we summarize the results.
The $\mathbb{Z}_n$ orbifold actions preserving ${\cal N}=6$ SUSY are
\beq
(n_1,n_2,n_3,n_4) = (n,n,n,n).
\eeq
The $\mathbb{Z}_n$ orbifold actions preserving ${\cal N}=4$ SUSY are
\beq
(n_1,n_2,n_3,n_4) = (n,n,\infty,\infty), \,\, (n,n,m,m), \,\,  (2,-2,1,\infty ) 
\eeq
The orbifold actions preserving ${\cal N}=2$ SUSY are infinitely many. For example, 
\beq
(n_1,n_2,n_3,n_4) = (n,-n,\infty,\infty), \,\, (2n,2n,n,\infty), \,\, (6n,3n,2n,\infty), \cr
\,\, (12n,4n,3n,\infty), \,\, \cdots , (n,n,n,3n), \cdots  
\eeq

\section{Moduli spaces of several orbifold gauge theories}

In this section, we study several examples of orbifold gauge theories which have
${\cal N}=4$ supersymmetry.
We consider the following three cases:
$$ (n_1,n_2,n_3,n_4) = (n,n,-n,-n), \,\, (n,n,\infty,\infty), \,\, (n,n,n,n).$$
We construct orbifold gauge theory corresponding to them, write quiver diagrams,
 and calculate moduli spaces of them.

\subsection{Orbifold gauge theory I \label {case I}}

In this section, we consider $\mathbb{Z}_n$ orbifold by the action 
$$ y^A \to e^{2\pi i/n_A} y^A$$
with
$$(n_1, n_2 ,n_3, n_4) = (n,n,-n,-n).$$
This is the special case for $(n,n,m,m)$ discussed in the previous section,
and is a subgroup of the $SU(4)_R$ symmetry.
As we will discuss later, 
we assume that Chern-Simons coupling $k$ before orbifolding
is quantized by $n$ as $k=k'n$ where $k' \in \mathbf Z$
in order to construct a consistent orbifold gauge theory.
As this $\mathbb{Z}_n$ action commutes with $SU(2) \times SU(2) \sim SO(4)$ symmetry,
which is a subgroup of the $SU(4)_R$ symmetry, 
this theory indeed has ${\cal N}=4$ supersymmetry.

According to \siki{def_y}, we put 
\begin{eqnarray}
y^1=Z^1, \quad y^2=Z^2, \quad y^3=W^{\dagger 1}, \quad y^4=W^{\dagger 2}.
\label{label_y}
\end{eqnarray}
Then, this orbifold action can be rewritten as 
$$Z^1 \to e^{2\pi i/n} Z^1, \qquad Z^2 \to e^{2\pi i/n} Z^2, \qquad
W^{\dagger 1} \to e^{-2\pi i/n} W^{\dagger 1} , \qquad 
W^{\dagger 2} \to e^{-2\pi i/n} W^{\dagger 2}.$$
Taking account that $\psi_A$ are transformed by the $SU(4)_R$ transformation
in the opposite way as $y^A$,
we can also write this action on the fermions using \siki{def_psi} as
$$\zeta^2 \to e^{-2\pi i/n} \zeta^2, \qquad 
\zeta^1 \to e^{-2\pi i/n} \zeta^1, \qquad
\omega^{\dagger 2} \to e^{2\pi i/n} \omega^{\dagger 2}, \qquad 
\omega^{\dagger 1} \to e^{2\pi i/n} \omega^{\dagger 1}.$$
We note that we cannot write this $\mathbb{Z}_n$ action in terms of superfields
because this $\mathbb{Z}_n$ action is not compatible with supersymmetry 
which is manifest in the superspace formalism in this notation.
Actually, if we put $y^1=Z^1$, $y^2=W^{\dagger 1}$, $y^3=Z^2$, $y^4=W^{\dagger 2}$
instead of \siki{label_y},
we could write this $\mathbb{Z}_n$ action 
in terms of superfields \siki{superfield} as
${\cal Z}^1 \to e^{2\pi i/n} {\cal Z}^1$, 
${\cal Z}^2 \to e^{-2\pi i/n} {\cal Z}^2$,  
${\cal W}^1 \to e^{-2\pi i/n} {\cal W}^1$, 
${\cal W}^2 \to e^{2\pi i/n} {\cal W}^2.$
However, in order for simplicity of the calculation of the moduli space,
we take the convention \siki{label_y}.

In order to construct orbifold gauge theory by $\mathbb{Z}_n$, 
we introduce $nN$ M2 branes, which means that $Z^A$ and $W^A$ are now $nN \times nN$ matrix.
Then, we impose the condition that $n$ sets of $N$ M2 branes 
are put at the position of mirror image each other.
Conditions imposed on $Z^A$ and $W^A$ due to this orbifolding are \cite{Douglas_Moore,Fuji,Klebanov}
\begin{eqnarray}
&Z^A = e^{2\pi i/n} \Omega Z^A \Omega^{\dagger} , \quad
W^A = e^{2\pi i/n} \Omega W^A \Omega^{\dagger} , \quad \cr
&\zeta^A =  e^{-2\pi i/n} \Omega \zeta^A \Omega^{\dagger} , \quad 
\omega^A =  e^{-2\pi i/n} \Omega \omega^A \Omega^{\dagger} , \quad 
\cr
&A^{\mu}= \Omega A^{\mu} \Omega^{\dagger} , \quad
\hat{A}^{\mu} = \Omega \hat{A}^{\mu} \Omega^{\dagger},
\label{DM_cond}
\end{eqnarray}
where $\Omega$ is defined as
$$ \Omega = \mathrm{diag} ({\mathbf 1}_{N \times N} , \,\, e^{2\pi i/n} {\mathbf 1}_{N \times N},  \,\,
e^{4\pi i/n} {\mathbf 1}_{N \times N},  \,\, \cdots   \,\, e^{2\pi i(n-1)/n} {\mathbf 1}_{N \times N}) .$$
This condition indicates that when $n$ sets of $N$ M2 branes 
are exchanged by $\Omega$, the phase factor arises, 
which is exactly the condition for mirror image.

This method of constructing orbifold gauge theory is known to work quite well for world volume theory of 
D-branes \cite{Douglas_Moore}, but is not justified $a$ $priori$ for that of M2 branes.
However, taking account that orbifold theory of M2 branes should reproduce that of D2 branes
at the place far from the orbifold fixed point in the moduli space,
it is plausible to assume that this method is applicable also for the M2 brane case
as performed in \cite{Fuji}, where consistent remaining supersymmetry and the D2 brane limit 
were observed.
Indeed, we will see later that moduli space of the theory 
constructed by this method is consistent with the assumption that 
this theory is the world volume theory of the M2 branes probing the orbifolded space
\footnote{There is some subtlety when $k'$ and $n$ are not coprime, as we will see later.}.

\begin{figure}
\centering
\input{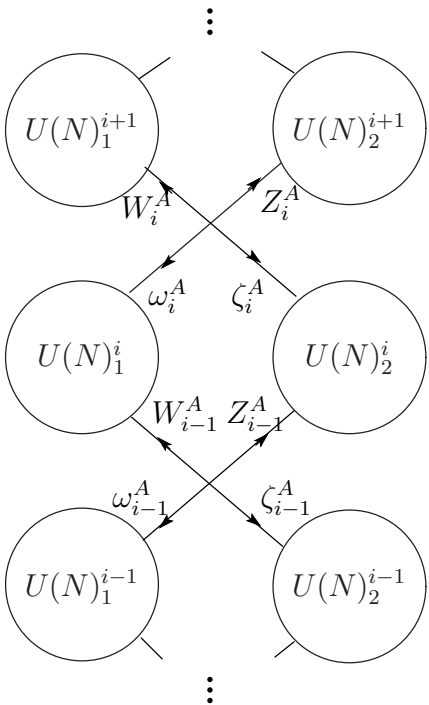}
\vspace{5mm}
\caption{Quiver diagram I}
\label{nn-n-n_Quiver}
\end{figure}

Solving the conditions (\ref{DM_cond}), we have
\begin{eqnarray}
&& \!\!\!\!\!\!\!\! 
Z^A = 
\left(
\begin{array}{cccccccccc}
0 & Z_1^A & \cr
 & 0 & Z_2^A \cr
 & & \ddots & \ddots \cr
 & & & 0 & Z_{n-1}^A \cr
Z_n^A & & & & 0
\end{array}
\right) , \quad 
W^A = 
\left(
\begin{array}{cccccccccc}
0 & W_1^A & \cr
 & 0 & W_2^A \cr
 & & \ddots & \ddots \cr
 & & & 0 & W_{n-1}^A \cr
W_n^A & & & & 0
\end{array}
\right) \cr 
&& \!\!\!\!\!\!\!\! 
\zeta^A = 
\left(
\begin{array}{cccccccccc}
0 & & & & \zeta_n^A &  \cr
\zeta_1^A & 0 &  \cr
 & \zeta_1^A & 0 & \cr
 & & \ddots & \ddots  \cr
 & & & \zeta_{n-1}^A& 0
\end{array}
\right) , \qquad 
\omega^A = 
\left(
\begin{array}{cccccccccc}
0 & & & & \omega_n^A &  \cr
\omega_1^A & 0 &  \cr
 & \omega_1^A & 0 & \cr
 & & \ddots & \ddots  \cr
 & & & \omega_{n-1}^A& 0
\end{array}
\right) \cr \cr
&& \!\!\!\!\!\!\!\! 
A^{\mu} = \mathrm{diag} (A_1^{\mu} , A_2^{\mu}, \cdots A_n^{\mu}) , \qquad \qquad \qquad
\hat{A}^{\mu} = \mathrm{diag} (\hat{A}_1^{\mu} , \hat{A}_2^{\mu}, \cdots \hat{A}_n^{\mu})
\label{sol_orb}
\end{eqnarray}
Gauge symmetries and matter contents are conveniently summarized by quiver diagram
as in Figure \ref{nn-n-n_Quiver}.
The situation is different whether $n$ is even or odd.
When $n$ is even, $\mathbb{Z}_2$ subgroup, which invert the sign of the matter fields,
of the $\mathbb{Z}_n$ orbifold action are already included in the 
$\mathbb{Z}_k$ action of the original ABJM theory because $k=k'n$ as we will see later.
In other words, the orbifold action which we are considering are redundant,
and thus, the quiver gauge theory constructed above are not good description.
Reflecting this fact, the quiver diagram divides into two parts when $n$ is even.
Similar situation occurs in the orbifold gauge theory for usual D-brane
when we formally impose the constraint of orbifold to the theory 
already orbifolded by the same discrete action.

It does not happen when $n$ is odd.
In the following, we concentrate on the odd case.

Here, the important point is that the action have to be multiplied by $1/n$
\footnote{We thank Koji Hashimoto for the discussion on this point.},
which is explained as follows.
For example, in the branch 3 in (\ref{phase_3}), where the M2 brane
is far away from the orbifold singularity,
 the gauge groups are 
spontaneously broken from $\left( U(1) \times U(1) \right)^n$
into $U(1)_{\tilde{b}}$, under which no matter fields are charged.
By expanding the fields around this point in the moduli space,
and substituting these into the original unorbifolded action,
we obtain the usual $U(1)$ action multiplied by $n$ \cite{Mukhi_Papageorgakis}.
In order to reproduce correct membrane tension in this phase,
the action have to be multiplied by $1/n$ from the beginning; 
\begin{eqnarray}
S = \frac{1}{n} \int d^3 x \left[ \frac{k}{4 \pi} \varepsilon^{\mu\nu\lambda} \mathrm{Tr} 
\left(
A_{\mu} \partial _{\nu} A_\lambda + \frac{2i}{3} A_{\mu} A_{\nu} A_{\lambda} \right)
+ \cdots \right]
\end{eqnarray}
Thus, the Chern-Simons coupling for each gauge group is $k'=k/n$,
where $k$ is the original Chern-Simons coupling before the orbifolding.
In order to construct a consistent orbifold gauge theory,
$k'$ must be an integer, which means that $k$ must be quantized by $n$.
Therefore, our orbifolded gauge theory will represent M2 branes on
$(\mathbb{C}^4/\mathbb{Z}_k)/\mathbb{Z}_n =\mathbb{C}^4/(\mathbb{Z}_{k'n}\times \mathbb{Z}_n) $.


From here, we calculate the moduli space of this theory.
The bosonic potential of 
the ABJM theory consists of the one originates from $F$ term, which we denote $V_F$,
and the one originates from $D$ term, which we denote $V_D$.
They are rewritten as 
\begin{eqnarray}
V_D &=& \frac{4\pi^2}{k^2} \mathrm{Tr} \left[ 
 (Z^A Z_A^{\dagger} - W^{\dagger A} W_A) Z^B
- Z^B ( Z_A^{\dagger} Z^A - W_A W^{\dagger A} ) \right] \cr
&&\qquad \qquad \left[ 
 (Z^C Z_C^{\dagger} - W^{\dagger C} W_C) Z^B
- Z^B ( Z_C^{\dagger} Z^C - W_C W^{\dagger C} ) \right]^{\dagger} \cr
&&+ \frac{4\pi^2}{k^2} \mathrm{Tr} \left[ 
 (W^A W_A^{\dagger} - Z^{\dagger A} Z_A) W^B
- W^B ( W_A^{\dagger} W^A - Z_A Z^{\dagger A} ) \right] \cr
&&\qquad \qquad \left[ 
 (W^C W_C^{\dagger} - Z^{\dagger C} Z_C) W^B
- W^B ( W_C^{\dagger} W^C - Z_C Z^{\dagger C} ) \right]^{\dagger} \cr
V_F &=& - \frac{16 \pi^2}{k^2} \mathrm{Tr} 
\left[ W_A Z^B W_C - W_C Z^B W_A \right] 
\left[ W_A Z^B W_C - W_C Z^B W_A \right] ^{\dagger} \cr
&&- \frac{16\pi^2}{k^2} \mathrm{Tr} 
\left[ Z^A W_B Z^C - Z^C W_B Z^A \right] 
\left[ Z^A W_B Z^C - Z^C W_B Z^A \right] ^{\dagger}
\end{eqnarray}
The conditions for supersymmetric vacua are $V_D=0$ and $V_F=0$,
The former are little bit different form the usual form of $D$ term condition and can be simplified as 
\begin{eqnarray}
(Z^A Z_A^{\dagger} - W^{\dagger A} W_A) Z^B
- Z^B ( Z_A^{\dagger} Z^A - W_A W^{\dagger A}) = 0 \cr
(W^A W_A^{\dagger} - Z^{\dagger A} Z_A) W^B
- W^B ( W_A^{\dagger} W^A - Z_A Z^{\dagger A}) = 0
\label{general_D_term}
\end{eqnarray}
On the other hand, the latter can be rewritten in the same form as the usual
$F$ term conditions as
\begin{eqnarray}
W_A Z^B W_C - W_C Z^B W_A = 0 \cr
Z^A W_B Z^C - Z^C W_B Z^A = 0 ,
\label{general_F_term}
\end{eqnarray}

By substituting (\ref{sol_orb}) into \siki{general_D_term} and \siki{general_F_term},
 we calculate the moduli space of this theory.
The condition \siki{general_D_term} from the $D$ term is given by
\begin{eqnarray}
&&\!\!\!\!\!\!\!\!\!\! (Z^{A}_i Z_{iA}^{\dagger} - W^{\dagger A}_{i-1} W_{i-1 A}) Z^{B}_i
= Z^{B}_i ( Z_{iA}^{\dagger} Z^{A}_i - W_{i+1 A} W_{i+1}^{ \dagger A}) \cr
&&\!\!\!\!\!\!\!\!\!\! (W^{A}_i W_{iA}^{\dagger} - Z_{i-1}^{\dagger A} Z_{i-1A}) W^{B}_i
= W^{B}_i ( W_{iA}^{\dagger} W_i^{A} - Z_{i+1 A} Z^{\dagger A}_{i+1}) 
 \quad (i=1,\cdots n) 
\end{eqnarray}
while the $F$ term condition \siki{general_F_term} is given by
\begin{eqnarray}
&& Z^A_{i-1} W^B_{i} Z^C_{i+1} = Z^C_{i-1} W^B_{i} Z^A_{i+1}  \cr
&& W^A_{i-1} Z^B_{i} W^C_{i+1} = W^C_{i-1} Z^B_{i} W^A_{i+1}
, \qquad (i=1, \cdots n)
\end{eqnarray}

In the following, we limit the Abelian case, that is, $N=1$.
In this case, these conditions are simplified as follows.
The conditions from the $D$ term are given by
\begin{eqnarray}
&&(W^{\dagger A}_{i-1} W_{i-1 A} - W_{i+1 A} W_{i+1}^{\dagger A}) Z^{B}_i = 0, \cr
&&(Z_{i-1}^{\dagger A} Z_{i-1 A} - Z_{i+1 A} Z_{i+1}^{\dagger A}) W^{B}_i = 0.
 \quad (i=1,\cdots n) \label{D_term}
\end{eqnarray}
while $F$ term conditions are given by
\begin{eqnarray}
&& W^B_{i} ( Z^A_{i-1} Z^C_{i+1} - Z^A_{i+1} Z^C_{i-1} ) = 0 ,\cr
&& Z^B_{i} ( W^A_{i-1} W^C_{i+1} - W^A_{i+1} W^C_{i-1} ) = 0.
 \qquad (i=1, \cdots n)
\label{F_term}
\end{eqnarray}

Actually, solutions of these equations have various branches like

\begin{eqnarray}
&&1. \qquad Z=0, \quad W: {\rm arbitrary} \cr
&&2. \qquad W=0, \quad Z: {\rm arbitrary} \cr
&&3. \qquad W_{i-1}^{\dagger A} W_{i-1 A} = W_{i+1 A} W_{i+1}^{\dagger A}, \quad
Z_{i-1}^{\dagger A} Z_{i-1 A} = Z_{i+1 A} Z_{i+1}^{\dagger A} \cr
&&\qquad\quad W_{i-1}^A W_{i+1}^B = W_{i+1}^A W_{i-1}^B, \quad
Z_{i-1}^A Z_{i+1}^B = Z_{i+1}^A Z_{i-1}^B
\label{phase_3}
\end{eqnarray}

Existence of the first and the second branch is quite natural.
Remember that there are two kinds of orbifold action;
the $\mathbb{Z}_k$ action%
\footnote{As we will see later, we have to assume that $k'$ and $n$
are coprime, where $k=nk'$, in order that this $\mathbb{Z}_k$ action are 
reflected to the moduli space.}, 
which has already acted on the original ABJM theory,
and the $\mathbb{Z}_n$ action, which we are now considering.
The points satisfying $Z=0$ are orbifold singularities
because the combination $k' \mathbb{Z}_{k} + \mathbb{Z}_n$
acts trivially on such region, and thus, the first branch appears there.
In this branch, the moduli space will be an orbifold of $(\mathbb{C}^2)^n$ 
because $W$ is arbitrary.
This implies that there are $n$ objects with $T_{M2}/n$ tension,
where $T_{M2}$ is a M2 brane tension.  
In the usual D-brane case, corresponding degrees of freedom appears 
at an orbifold fixed point as fractional D-branes \cite{fractional}.
Thus, also in this case, 
we could interpret that it is due to the existence of ``fractional membranes''.

In the following, we concentrate on the third branch .
For solving the equations for $Z$, we put 
\begin{eqnarray}
Z_{i}^A = r_{i}^A e^{i \theta_{i}^A}
\end{eqnarray}
and substitute this into the above equations.
Then, we obtain
\begin{eqnarray}
r_{i-1}^1 r_{i+1}^2 = r_{i-1}^2 r_{i+1}^1 &\to& 
\frac{r_{i-1}^1}{r_{i-1}^2} = \frac{r_{i+1}^1}{ r_{i+1}^2} \equiv c \cr
e^{i (\theta_{i-1}^1 + \theta_{i+1}^2 ) } 
= e^{i (\theta_{i-1}^2 + \theta_{i+1}^1) } 
&\to& \theta_{i-1}^1 + \theta_{i+1}^2 = \theta_{i-1}^2 + \theta_{i+1}^1 
,\quad (\mathrm{mod}\,\, 2\pi) \cr
&\to& \theta_{i+1}^2 - \theta_{i+1}^1 =  \theta_{i-1}^2 - \theta_{i-1}^1 \equiv \theta \cr
(r_{i-1}^1)^2 + (r_{i-1}^2)^2 = (r_{i+1}^1)^2 + (r_{i+1}^2)^2 &\to& 
(1+c^2) (r_{i-1}^2)^2 = (1+c^2) (r_{i+1}^2)^2 \cr
&\to& r_{i-1}^A = r_{i+1}^A \equiv r^A
\end{eqnarray}

Especially, for $n$ odd, we have
\begin{eqnarray}
r_{i}^A = r^A , \qquad \theta_{i}^2 - \theta_{i}^1 = \theta,
\label{const1}
\end{eqnarray}
for arbitrary $i$.
The analysis for $W$ is the same as this.
We obtain the solution
\begin{eqnarray}
Z_{i}^1 = r^1 e^{i \theta_i} , \qquad Z_{i}^2 = r^2 e^{i (\theta_i + \theta)}, \qquad
W_{i}^1 = R^1 e^{i \phi_i} , \qquad W_{i}^2 = R^2 e^{i (\phi_i + \phi)} .
\label{result1}
\end{eqnarray}

At this stage, we do not consider the gauge symmetry.
In order to obtain the moduli space, 
we consider the remaining gauge symmetry, which is a subgroup of $U(n)_1 \times U(n)_2$.
The remaining gauge symmetry transformation 
should compatible with the orbifold action $\mathbb{Z}_n$.
The $U \in U(n)_1$ and $\hat{U} \in U(n)_2$ 
commute with the orbifold action
if they satisfy the following conditions
\begin{eqnarray}
\Omega^{-1} U \Omega = a U , \qquad \Omega^{-1} \hat{U} \Omega = a \hat{U},
\end{eqnarray}
where $a$ is an arbitrary phase factor.
Due to this phase factor, the remaining gauge symmetries which 
are compatible with the orbifold action 
are both $(U(1) \times U(1))^n$ given in Figure \ref{nn-n-n_Quiver}
and the following global discrete symmetry generated by
$$ U = \hat{U} = S,$$
where 
\begin{eqnarray}
S= 
\left(
\begin{array}{cccccccccc}
0 & 1 & \cr
 & 0 & 1 \cr
 & & \ddots & \ddots \cr
 & & & 0 & 1 \cr
1 & & & & 0
\end{array}
\right).
\end{eqnarray}
This transformation changes the indices of
$U(1)_1^i \times U(1)_2^i$ into those of $U(1)_1^{i+1} \times U(1)_2^{i+1}$
and act on each field as
\begin{eqnarray}
&&Z^A \to S Z^A S^{\dagger} = 
\left(
\begin{array}{cccccccccc}
0 & Z_2^A & \cr
 & 0 & Z_3^A \cr
 & & \ddots & \ddots \cr
 & & & 0 & Z_n^A \cr
Z_1^A & & & & 0
\end{array}
\right) \cr
&&W^A \to S W^A S^{\dagger} = 
\left(
\begin{array}{cccccccccc}
0 & W_2^A & \cr
 & 0 & W_3^A \cr
 & & \ddots & \ddots \cr
 & & & 0 & W_n^A \cr
W_1^A & & & & 0
\end{array}
\right) .
\end{eqnarray}
We call this $\mathbb{Z}_n$ symmetry shift symmetry.

At this stage, seemingly remaining gauge symmetry is $(U(1)\times U(1))^n$ and the shift symmetry.
Although actual gauge symmetry is 
a subgroup of this symmetry due to the quantization 
condition of the Chern-Simons term,
the invariant combination under $(U(1)\times U(1))^n$ and the shift symmetry 
becomes at least moduli parameter.
Actual moduli space is larger than the space parametrized by 
the parameters.

At this stage, we can say that the following seven are moduli parameters.
\begin{eqnarray}
&&r^1 \equiv |Z_i^{A=1}|, \qquad
r^2 \equiv |Z_i^{A=2}|, \qquad
R^1 \equiv |W_i^{A=1}|, \qquad
R^2 \equiv |W_i^{A=2}|, \cr \cr
&&e^{i\theta} \equiv \left. \frac{Z_i^{A=2}}{Z_i^{A=1}} \right/ 
\left| \frac{Z_i^{A=2}}{Z_i^{A=1}} \right| ,\qquad 
e^{i\phi} \equiv \left. \frac{W_i^{A=2}}{W_i^{A=1}} \right/ 
\left| \frac{W_i^{A=2}}{W_i^{A=1}} \right| ,\cr \cr
&&e^{i\sum_j (\theta^j + \phi^j)} \equiv \left.
\displaystyle \prod _{j=1}^n Z_j^{A=1} W_j^{B=1} \right/
\left| \displaystyle \prod _{j=1}^n Z_j^{A=1} W_j^{B=1} \right|
\label{moduli_param1}
\end{eqnarray}


In \cite{ABJM}, the Higgsing of $U(N)\times U(N)$
to $\prod_i U(1)^i_1 \times U(1)^i_2$ in the ABJM theory 
was considered.
In this case, (anti-)bifundamental matters $Z^i$ and $W^i$ are 
charged under $U(1)^i_1 \times U(1)^i_2$.
For each $i$, no matters are charged under $U(1)^i_1+U(1)^i_2$
and the corresponding gauge fields couple only through the Chern-Simons term. 
Thus, we can dualize these $N$ gauge fields and 
the corresponding $N$ constraints appear.

Situation is different in our model.
In this case, (anti-)bifundamental matters $Z^i$ and $W^i$ are charged under $U(1)^i_1 \times U(1)^{i+1}_2$
and $U(1)^{i+1}_1 \times U(1)^{i}_2$, respectively.
Thus, only 
$\sum_i U(1)^i_1 + U(1)^i_2$, which we denote $U(1)_{\tilde{b}}$,
is 
a combination of symmetries under which no matter fields are charged.
The corresponding combination of gauge fields
\begin{eqnarray}
A^{\mu}_{\tilde{b}} = \mathrm{Tr} (A^{\mu}_{U(n)_1} + \hat{A}^{\mu}_{U(n)_2})
= \displaystyle \sum_i \left( A^{\mu}_{1 i} + \hat{A}^{\mu}_{2 i} \right) 
\label{def_Abt}
\end{eqnarray}
couples through Chern-Simons term with
\begin{eqnarray}
A^{\mu}_b = \mathrm{Tr} (A^{\mu}_{U(n)_1} - \hat{A}^{\mu}_{U(n)_2})
= \displaystyle \sum_i \left( A^{\mu}_{1 i} - \hat{A}^{\mu}_{2 i} \right). 
\label{Ab}
\end{eqnarray}
Actually, 
invariant combinations under the shift symmetry are 
only $U(1)_{b}$ and $U(1)_{\tilde{b}}$.
Thus it is natural to impose quantization condition only to this invariant combination.

Because $A^{\mu}_{\tilde{b}}$ is invariant under the shift symmetry and does not couple to
matter fields, we can dualize this field.
That is, by adding 
$$S= \frac{1}{2\pi} \int \tau(x) \varepsilon_{\mu\nu\rho} \partial^{\mu}F^{\nu\rho}_{\tilde{b}},$$
where $\tau(x)$ is a Lagrange multiplier,
we can regard $F_{\tilde{b}}$ as 
an elementary field instead of $A_{\tilde{b}}$.
Here, we see that $\tau(x)$ has to be periodic as
$$\tau(x) \sim \tau(x)+2\pi$$
from the quantization condition 
$$\frac{1}{2}\int \varepsilon_{\mu\nu\rho} \partial^{\mu}F^{\nu\rho}_{\tilde{b}}
= \int_M dF_{\tilde{b}} = \int_{\partial M} F_{\tilde{b}} = 2\pi \mathbb{Z}.$$
Normalization of the last equality is followed by the normalization of \siki{def_Abt}.

As $F_{\tilde{b}}$ is an auxiliary field, we can integrate it out.
As an equation of motion for $F_{\tilde{b}}$ is
$$A_{b}^{\mu}(x) = \frac{1}{k'} \partial^{\mu} \tau(x),$$
when the gauge transformation act on gauge field as
$$A_{b}^{\mu}(x) \to A_{b}^{\mu}(x) + \partial^{\mu} \alpha (x),$$
$\tau (x)$ is transformed as
$$\tau(x) \to \tau(x) + k' \alpha(x).$$
With the normalization of \siki{Ab}, we find that $\alpha(x) \sim \alpha(x) + 2\pi$.
Thus, when we gauge fix as $\tau(x)=0$,
the gauge transformation 
$$\alpha(x)=\frac{2 \pi m}{k'}$$
still remains.
From the above discussion, we found that $U(1)_{b}$ symmetry is broken to $\mathbb{Z}_k$ symmetry
\begin{eqnarray}
&&\!\!\!\!\!\!\!\!\!\!\!\!\!\!\!\!\!\!
 (\theta_1,\theta_2, \cdots \theta_n, \phi_1, \phi_2, \cdots \phi_n)  \cr \cr
&& \!\!\!\! \to
\left( \theta_1 + \frac{2\pi}{k'}, \theta_2 + \frac{2\pi}{k'}, \cdots, \theta_n + \frac{2\pi}{k'},
\phi_1 - \frac{2\pi}{k'}, \phi_2 - \frac{2\pi}{k'}, \cdots, \phi_n - \frac{2\pi}{k'} \right). 
\label{div_k} 
\end{eqnarray}

On the other hand, traceless $U(1)^{2(n-1)}$ symmetries, which is included in 
$SU(n) \times SU(n) \subset U(n) \times U(n)$
still remains as a gauge symmetry.

If the discussion above is justified, a new moduli parameter  
other than (\ref{moduli_param1}) is 
$$e^{i \sum_j (\theta_j - \phi_j)}$$
but under the equivalence relation (\ref{div_k}).
Combined with a moduli parameter $e^{i \sum_j (\theta_j + \phi_j)}$ in (\ref{moduli_param1}),
we can regard 
\begin{eqnarray}
e^{i \sum_j \theta_j} \equiv \left.
\displaystyle \prod _{j=1}^n Z_j^{A=1} \right/
\left| \displaystyle \prod _{j=1}^n Z_j^{A=1} \right| \cr
e^{i \sum_j \phi_j} \equiv \left.
\displaystyle \prod _{j=1}^n W_j^{B=1} \right/
\left| \displaystyle \prod _{j=1}^n W_j^{B=1} \right|
\end{eqnarray}
as basis of moduli parameters.

In summary, moduli parameters at this stage are
\begin{eqnarray}
r^1 , \quad
r^2 , \quad
R^1 , \quad
R^2 , \quad
e^{i\theta} ,\quad 
e^{i\phi} ,\quad
e^{i\Theta} , \quad
e^{i\Phi}
\label{moduli_param_I}
\end{eqnarray}
under the equivalence relation (\ref{div_k}),
where we put $\Theta=\sum_j \theta_j$ and $\Phi=\sum_j \phi_j$.

Considering the metric of the moduli space 
in terms of these coordinates \siki{moduli_param_I}, we obtain
\begin{eqnarray}
ds^2 &=& (dr^1)^2 + (dr^2)^2 + (dR^1)^2 + (dR^2)^2 \cr
&&+ (r^1)^2 \left[ d \left( \frac{\Theta}{n} \right) \right] ^2 
+ (r^2)^2 \left[ d \left( \frac{\Theta}{n} + \theta \right) \right] ^2 \cr
&&+ (R^1)^2 \left[ d \left( \frac{\Phi}{n} \right) \right] ^2 
+ (r^2)^2 \left[ d \left( \frac{\Phi}{n} + \phi \right) \right] ^2,
\end{eqnarray}
where we restricted the scalar kinetic term
to the subspace defined by (\ref{const1}) 
and gauged away the $(U(1) \times U(1))^{n-1}$ parts which 
are parametrized by $\theta_i-\theta_{i+1}$ and $\phi_i-\phi_{i+1}$.
Thus, we find $\mathbb{C}_n \times \mathbb{C}_n$ orbifold structure.
The moduli space is parametrized by following coordinates;
\begin{eqnarray}
&&Z^1= r^1 \exp \left( i\frac{\Theta}{n} \right) , \qquad
Z^2= r^2 \exp \left( i\frac{\Theta}{n} + \theta \right) ,\cr
&&W^1= R^1 \exp \left( i\frac{\Phi}{n} \right) , \qquad
W^2= R^2 \exp \left( i\frac{\Phi}{n} + \phi \right) ,
\label{moduli_I}
\end{eqnarray}
where $\Theta \sim \Theta +2 \pi$ and $\Phi \sim \Phi + 2\pi$.
Indeed, the moduli parameters are identified
by the $\mathbb{C}_n \times \mathbb{C}_n$ 
which transform $Z^A$ and $W^A$ independently.
Changing the basis of this discrete action,
we can see that the moduli parameters are identified by the following actions; 
\begin{eqnarray}
\left( Z^A ,  \,\, W^A \right) \to \left( e^{2\pi i/n} Z^A , \,\, e^{-2\pi i/n} W^A \right)
\label{inv_1} \\
\left( Z^A ,  \,\, W^A \right) \to \left( e^{2\pi i/n} Z^A , \,\, e^{2\pi i/n} W^A \right).
\label{inv_2}
\end{eqnarray}

Rewriting the identifications (\ref{div_k}), (\ref{inv_1}), and (\ref{inv_2})
 in terms of the complex coordinates $y^A$,
we have
\begin{eqnarray}
&&(y^1,y^2,y^3,y^4) \to (e^{\frac{2\pi i}{k'}}y^1 ,e^{\frac{2\pi i}{k'}}y^2 
,e^{\frac{2\pi i}{k'}}y^3 ,e^{\frac{2\pi i}{k'}}y^4 ) , \label{Z_k'}\\
&&(y^1,y^2,y^3,y^4) \to (e^{\frac{2\pi i}{n}}y^1 ,e^{\frac{2\pi i}{n}}y^2 
,e^{\frac{2\pi i}{n}}y^3 ,e^{\frac{2\pi i}{n}}y^4 ) , \label{Z_n_1}\\
&&(y^1,y^2,y^3,y^4) \to (e^{\frac{2\pi i}{n}}y^1 ,e^{\frac{2\pi i}{n}}y^2 
,e^{\frac{-2\pi i}{n}}y^3 ,e^{\frac{-2\pi i}{n}}y^4 ). \label{Z_n_2}
\end{eqnarray}
Thus, the moduli space is 
$\mathbb{C}^4/\left( \mathbb{Z}_{k'} \times \mathbb{Z}_n \times \mathbb{Z}_n \right)$.

As discussed previously, in order to construct a consistent orbifold gauge theory,
we have to assume that $k$ is quantized by $n$,
which plays a crucial role to the analysis of the moduli space.
We see that the $\mathbb{Z}_{k'}$ action (\ref{Z_k'}), where $k=k'n$, and 
the $\mathbb{Z}_n$ action \siki{Z_n_1} are subgroups of the original $\mathbb{Z}_k$ action.
Especially, if we assume that $k'$ and $n$ are coprime, 
the original $\mathbb{Z}_k$ action can be decomposed into these two discrete groups as
 $\mathbb{Z}_{k'} \times \mathbb{Z}_n$.
On the other hand, the $\mathbb{Z}_n$ action \siki{Z_n_2} reflects the 
$\mathbb{Z}_n$ orbifold action which we are considering.
Thus, if $k'$ and $n$ are coprime,
we find that the moduli space of this theory is 
consistent with the fact that the M2-brane probes 
$\mathbb{C}^4/(\mathbb{Z}_k \times \mathbb{Z}_n)$,
where $\mathbb{Z}_k$ act complex coordinates $y^A$ as  
$$
(y^1,y^2,y^3,y^4) \to (e^{\frac{2\pi i}{n}}y^1 ,e^{\frac{2\pi i}{n}}y^2 
,e^{\frac{-2\pi i}{n}}y^3 ,e^{\frac{-2\pi i}{n}}y^4 ).
$$
while $\mathbb{Z}_k$ act as
$$
(y^1,y^2,y^3,y^4) \to (e^{\frac{2\pi i}{k}}y^1 ,e^{\frac{2\pi i}{k}}y^2 
,e^{\frac{2\pi i}{k}}y^3 ,e^{\frac{2\pi i}{k}}y^4 ).
$$
This is the expected results.

However, if $k'$ and $n$ are not coprime, we could not obtain the expected 
moduli space by our analysis.
It indicates the lack of our understanding of 
the orbifolding of the membrane action.


\subsection{Orbifold gauge theory II}

In this subsection, we consider $\mathbb{Z}_n$ orbifold by the action 
$$ y^A \to e^{2\pi i/n_A} y^A$$
with
$$(n_1 , n_2 ,n_3, n_4) = (n,n,\infty,\infty),$$
which is discussed in \cite{Klebanov}.
We can see as in the previous subsection that this theory has also ${\cal N}=4$ supersymmetry.
By using \siki{label_y}, 
we can write this action in terms of fields as%
\footnote{Similarly to the situation in the previous subsection,
we could write equivalent action in terms of superfields as
$$
{\cal Z}^1 \to e^{2\pi i/n} {\cal Z}^1 , \quad
{\cal W}^1 \to e^{-2\pi i/n} {\cal W}^1 , \quad
{\cal Z}^2 \to {\cal Z}^2, \quad
{\cal W}^2 \to {\cal W}^2,
$$
which we do not use this convention.}
\begin{eqnarray} Z^A \to e^{2\pi i/n} Z^A , \quad
W^A \to W^A , \quad
\zeta^A \to \zeta^A, \quad
\omega^A \to e^{-2\pi i/n} \omega^A.
\end{eqnarray}
This action is actually the combination of the action
$$ Z^A \to e^{2\pi i/2n} Z^A , \quad
W^A \to e^{-2\pi i/2n} W^A , \quad
\zeta^A \to e^{-2\pi i/2n} \zeta^A , \quad
\omega^A \to e^{2\pi i/2n} \omega^A ,$$
which is a subgroup of $SU(4)_R$ symmetry and the action
$$ Z^A \to e^{2\pi i/2n} Z^A , \quad
W^A \to e^{2\pi i/2n} W^A , \quad
\zeta^A \to e^{2\pi i/2n} \zeta^A , \quad
\omega^A \to e^{2\pi i/2n} \omega^A ,$$
which is a subgroup of $U(1)_{\tilde{b}}$ symmetry.

Conditions imposed on $Z^A$ and $W^B$ due to this orbifolding are
\begin{eqnarray}
&Z^A = e^{2\pi i/n} \Omega Z^A \Omega^{\dagger} , \quad
W^A = \Omega W^A \Omega^{\dagger} , \quad
\zeta^A = e^{- 2\pi i/n} \Omega \zeta^A \Omega^{\dagger} , \quad
\omega^A = \Omega \omega^A \Omega^{\dagger} \cr
&A^{\mu} = \Omega A^{\mu} \Omega^{\dagger} , \qquad
\hat{A}^{\mu} = \Omega \hat{A}^{\mu} \Omega^{\dagger}
\end{eqnarray}
Solving these conditions, we have
\begin{eqnarray}
&Z^A = 
\left(
\begin{array}{cccccccccc}
0 & Z_1^A & \cr
 & 0 & Z_2^A \cr
 & & \ddots & \ddots \cr
 & & & 0 & Z_{n-1}^A \cr
Z_n^A & & & & 0
\end{array}
\right) , \qquad 
W^A = \mathrm{diag} (W_1^A, W_2^A, \cdots, W_n^A) 
\cr \cr 
& \zeta^A = \mathrm{diag} (\zeta_1^A, \zeta_2^A, \cdots, \zeta_n^A)
, \qquad
\omega^A = 
\left(
\begin{array}{cccccccccc}
0 & & & & \omega_n^A \cr
\omega_1^A & 0 &  \cr
 & \omega_2^A & 0 & \cr
 & & \ddots & \ddots &  \cr
 & & & \omega_{n-1}^A & 0
\end{array}
\right) \cr \cr
&A^{\mu} = \mathrm{diag} (A^{\mu}_1 , A^{\mu}_2, \cdots A^{\mu}_n) , \qquad \qquad \qquad 
\hat{A}^{\mu} = \mathrm{diag} (\hat{A}^{\mu}_1 , \hat{A}^{\mu}_2, \cdots \hat{A}^{\mu}_n)
\label{sol_orb3}
\end{eqnarray}
Gauge symmetries and matter contents are conveniently summarized by quiver diagram
as in Figure \ref{nn00_Quiver}.
\begin{figure}
\centering
\input{quiver3.tpic}
\vspace{5mm}
\caption{Quiver diagram II}
\label{nn00_Quiver}
\end{figure}

In order to calculate moduli space, we solve (\ref{general_D_term}) and (\ref{general_F_term})
by substituting (\ref{sol_orb3}).
The conditions from the $D$ term \siki{general_D_term} are given by
\begin{eqnarray}
&&\!\!\!\!\!\!\!\!\!\! (Z^{A}_i Z_{iA}^{\dagger} - W_i^{\dagger A} W_{i A}) Z_i^{B}
= Z_i^{B} ( Z_{iA}^{\dagger} Z_i^{A} - W_{i+1 A} W_{i+1}^{\dagger A}) \cr
&&\!\!\!\!\!\!\!\!\!\! (W_i^{A} W_{iA}^{\dagger} - Z_{i-1}^{\dagger A} Z_{i-1 A}) W_i^{B}
= W_i^{B} ( W_{iA}^{\dagger} W_i^{A} - Z_{iA} Z_i^{\dagger A}) 
 \quad (i=1,\cdots n) 
\end{eqnarray}
while the $F$ term conditions (\ref{general_F_term}) are given by
\begin{eqnarray}
&& Z^A_{i} W^B_{i+1} Z^C_{i+1} = Z^C_{i} W^B_{i+1} Z^A_{i+1}  \cr
&& W^A_{i} Z^B_{i} W^C_{i+1} = W^C_{i} Z^B_{i} W^A_{i+1}
, \qquad (i=1, \cdots n)
\end{eqnarray}

Here, again, we concentrate on the Abelian case and
these two conditions are simplified as 
\begin{eqnarray}
&&Z_i^{B} ( W_{i+1 A} W_{i+1}^{\dagger A}- W_{iA} W_i^{\dagger A}) = 0 \cr
&&W_i^{B} ( Z_{iA} Z_i^{\dagger A} - Z_{i-1 A} Z_{i-1}^{\dagger A}) = 0  
 \quad (i=1,\cdots n) 
\end{eqnarray}
and
\begin{eqnarray}
&& W^{B}_{i+1} ( Z^A_{i} Z^C_{i+1} - Z^A_{i+1} Z^C_{i} ) = 0 \cr
&& Z^B_{i} ( W^A_{i} W^C_{i+1} - W^A_{i+1} W^C_{i} ) = 0
, \qquad (i=1, \cdots n)
\end{eqnarray}

Actually, solutions of these equations have various branches like
\begin{enumerate}
\item $Z=0, \quad W:$arbitrary 

\item $W=0, \quad Z:$arbitrary 

\item $W_{i}^A W_i^{\dagger A}  = W_{i+1 A} W_{i+1}^{\dagger A}, \quad
Z_{iA} Z_i^{\dagger A} = Z_{i+1 A} Z_{i+1}^{\dagger A}$ \\
$W_{i}^A W_{i+1}^B = W_{i+1}^A W_{i}^B, \quad
Z_{i}^A Z_{i+1}^B = Z_{i+1}^A Z_{i}^B$.
\end{enumerate}

We concentrate on the third branch.
We put 
\begin{eqnarray}
Z_{i}^A = r_{i}^A e^{i \theta_{i}^A}
\end{eqnarray}
and substitute this into the above equations. Then, we obtain
\begin{eqnarray}
r_{i}^1 r_{i+1}^2 = r_{i}^2 r_{i+1}^1 &\to& 
\frac{r_{i}^1}{r_{i}^2} = \frac{r_{i+1}^1}{ r_{i+1}^2} \equiv c \cr
e^{i (\theta_{i}^1 + \theta_{i+1}^2 ) } 
= e^{i (\theta_{i}^2 + \theta_{i+1}^1) } 
&\to& \theta_{i}^1 + \theta_{i+1}^2 = \theta_{i}^2 + \theta_{i+1}^1 
,\quad (\mathrm{mod}\,\, 2\pi) \cr
&\to& \theta_{i+1}^2 - \theta_{i+1}^1 =  \theta_{i}^2 - \theta_{i}^1 \equiv \theta \cr
(r_{i}^1)^2 + (r_{i}^2)^2 = (r_{i+1}^1)^2 + (r_{i+1}^2)^2 &\to& 
(1+c^2) (r_{i}^2)^2 = (1+c^2) (r_{i+1}^2)^2 \cr
&\to& r_{i}^A = r_{i+1}^A \equiv r^A
\end{eqnarray}
In this theory,
we obtain, regardless of the parity of $n$,
$$ r_{i}^A = r^A , \qquad \theta_{i}^2 - \theta_{i}^1 = \theta$$
for arbitrary $i$.
The analysis for $W$ is the same as this
and we find
$$ R_{i}^A = R^A , \qquad \phi_{i}^2 - \phi_{i}^1 = \phi \, .$$
Thus we obtain the solution
\begin{eqnarray}
Z_{i}^1 = r^1 e^{i \theta_i} , \qquad Z_{i}^2 = r^2 e^{i (\theta_i + \theta)}, \qquad
W_{i}^1 = R^1 e^{i \phi_i} , \qquad W_{i}^2 
= R^2 e^{i (\phi_i + \phi)}.
\label{result2}
\end{eqnarray}

In the following, we consider remaining gauge symmetry.
The existence of the global shift symmetry and the breakdown of 
$U(1)_b$ symmetry into $\mathbb{Z}_{k'}$ due to the quantization of the Chern-Simons term 
can be discussed similarly to the previous subsection.
Then, the moduli parameters are given by \siki{moduli_I}.
Assuming that $k'$ and $n$ are coprime,
we see, by changing the basis of the discrete action, 
that the moduli parameters are identified by the action
\begin{eqnarray}
&&\left( Z^A ,  \,\, W^A \right) \to \left( e^{2\pi i/n} Z^A , \,\, e^{-2\pi i/n} W^A \right) \\
&&\left( Z^A ,  \,\, W^A \right) \to \left( e^{2\pi i/n} Z^A , \,\, W^A \right).
\end{eqnarray}
Thus, together with the equivalence relation (\ref{div_k}),
we see that the moduli space is
$\mathbb{C}^4 / (\mathbb{Z}_k \times \mathbb{Z}_n)$, 
where $\mathbb{Z}_k$ act on complex coordinate $y^A$ as
$$
(y^1,y^2,y^3,y^4) \to (e^{\frac{2\pi i}{k}}y^1 ,e^{\frac{2\pi i}{k}}y^2 
,e^{\frac{2\pi i}{k}}y^3 ,e^{\frac{2\pi i}{k}}y^4 ).
$$
while $\mathbb{Z}_n$ act as 
$$
(y^1,y^2,y^3,y^4) \to (e^{\frac{2\pi i}{n}}y^1 ,e^{\frac{2\pi i}{n}}y^2 ,y^3 ,y^4 ).
$$
This is consistent with the orbifold action.

This result is also consistent with the result of \cite{Imamura},
in which it was discussed that the moduli space of this theory for $k'=1$ is 
$\mathbb{C}^4 / (\mathbb{Z}_n \times \mathbb{Z}_n)
=\left( \mathbb{C}^2/ \mathbb{Z}_n\right)^2$.

Actually, if $n$ is odd,
this theory is essentially the same as that of the case of 
``Orbifold gauge theory I'' under the condition that $k$ is quantized by $n$ and 
that $k'$ and $n$ are coprime, where $k=k'n$.
By changing the basis of the orbifold action, we see that 
the $\mathbb{Z}_n \times \mathbb{Z}_k 
= \mathbb{Z}_n \times \mathbb{Z}_n \times \mathbb{Z}_{k'}$ action of this theory 
is equivalent to that of ``Orbifold gauge theory I''.


\subsection{Orbifold gauge theory III ?}

\begin{figure}
\centering
\input{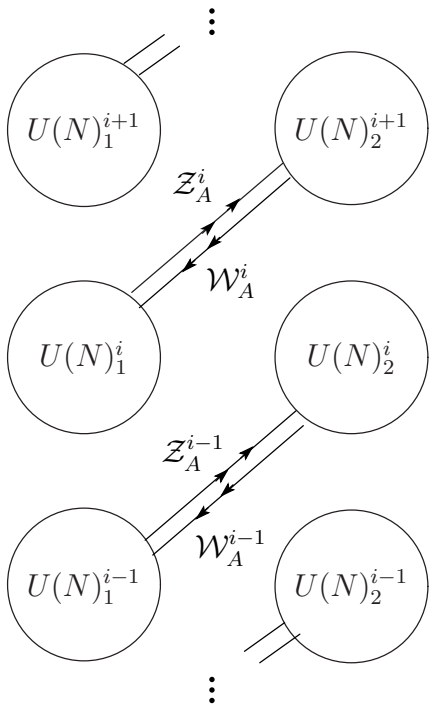}
\vspace{5mm}
\caption{Quiver diagram III}
\label{nnnn_Quiver}
\end{figure}

In this subsection, we consider the orbifold gauge theory by the $\mathbb{Z}_n$ action
\begin{eqnarray}
y^A \to e^{2\pi i/n} y^A. \label{orb_III}
\end{eqnarray}
Because this action commutes with $SU(4)_R$ symmetry, ${\cal N}=6$ supersymmetry preserves.
Rewritten in terms of superfields, this action becomes
$$ {\cal Z}^A \to e^{2\pi i/n} {\cal Z}^A ,\quad {\cal W}^A \to e^{-2\pi i/n} {\cal W}^A.$$
Similarly to the previous section, we impose the conditions 
\begin{eqnarray}
{\cal Z}^A = e^{2\pi i/n} \Omega {\cal Z}^A \Omega^{\dagger} , \qquad
{\cal W}^A = e^{-2\pi i/n} \Omega {\cal W}^A \Omega^{\dagger} , \qquad
{\cal V} = \Omega {\cal V} \Omega^{\dagger} , \qquad
\tilde{{\cal V}} = \Omega \tilde{{\cal V}} \Omega^{\dagger}.
\end{eqnarray}
Solving these equations, we obtain
\begin{eqnarray}
&{\cal Z}^A = 
\left(
\begin{array}{cccccccccc}
0 & {\cal Z}_1^A & \cr
 & 0 & {\cal Z}_2^A \cr
 & & \ddots & \ddots \cr
 & & & 0 & {\cal Z}_{n-1}^A \cr
{\cal Z}_n^A & & & & 0
\end{array}
\right) , \qquad
{\cal W}^A = 
\left(
\begin{array}{cccccccccc}
0 & & & & {\cal W}_n^A \cr
{\cal W}_1^A & 0 &  \cr
 & {\cal W}_2^A & 0 & \cr
 & & \ddots & \ddots &  \cr
 & & & {\cal W}_{n-1}^A & 0
\end{array}
\right) \cr \cr
&{\cal V} = \mathrm{diag} ({\cal V}_1 , {\cal V}_2, \cdots {\cal V}_n) , \qquad\qquad\qquad
\hat{{\cal V}} = \mathrm{diag} (\hat{{\cal V}}_1 , \hat{{\cal V}}_2, \cdots \hat{{\cal V}}_n)
\end{eqnarray}
The result described in the quiver diagram is Figure \ref{nnnn_Quiver}.
We obtain $n$ sets of decoupled original $U(N)\times U(N)$ theory.
Strictly speaking, they are not completely decoupled to each other 
but related only through the shift symmetry.
This is not expected results from the orbifold action.

As discussed in the previous section,
the Chern-Simons coupling $k$ before orbifolding 
have to be quantized by $n$.
This means that the left hand side and the right hand side 
of the orbifold action \siki{orb_III} are
already identified before orbifolding.
That is the reason why we obtained the unexpected results.
The situation is similar to the case where $n$ is even in ``Orbifold gauge theory I''.

\section{Conclusion}

In this paper, we have studied the 
$\mathbb{Z}_n$ orbifolds
of the ABJM theory with $U(N) \times U(N)$ gauge group.
Besides the models discussed in \cite{Klebanov}
we found, for example, the ${\cal N}=4$ supersymmetric 
membrane theory on $\mathbb{C}^4/
(\mathbb{Z}_k \times \mathbb{Z}_n)$.\footnote{
Recently, it was shown in \cite{Hosomichi2} that 
ABJM theory is actually a special case of the 
wide class of three dimensional ${\cal N}=4$ Chern-Simons theories
investigated in \cite{Hosomichi1}.
It is interesting to check that our orbifold gauge theories are 
also included in this class.}
The orbifold actions were taken as in
the orbifold of the D-brane world volume theory,
where the bi-fundamental representation of $U(N) \times U(N)$
are regarded as the adjoint representation of $U(N)$.

In order to verify the orbifold theories 
indeed describe the membranes on $\mathbb{C}^4/
(\mathbb{Z}_k \times \mathbb{Z}_n)$, 
we analyzed the moduli spaces of them.
Here, we had to assume that $k$ is given by $k=nk'$,
where $k'$ is integer, 
in order to construct a consistent orbifold gauge theory.
Together with the assumption that $n$ and $k'$ are coprime,
the moduli spaces were shown to agree with the orbifolded space.
However, if the $n$ and $k'$ are not coprime, we could not obtain an expected result.
Solving this problem will be an interesting future work.
It is also interesting to extend our work to
non-Abelian orbifolds.

\section*{Acknowledgments}
We would like to thank Hiroyuki Fuji for discussions and collaborations in early stages
of this project.
We thank Koji Hashimoto and Yosuke Imamura for discussion about the analysis of the moduli space.
S.~T.~is partly supported by the Japan Ministry of Education, Culture, Sports, Science and Technology. 
F.~Y.~is supported by JSPS fellowships for Young Scientists.

\end{document}